\documentclass[journal,12pt,onecolumn,draftclsnofoot]{IEEEtran}
\IEEEoverridecommandlockouts
\usepackage{algorithm}
\usepackage{algorithmicx}
\usepackage{algpseudocode}
\usepackage[latin1]{inputenc}
\usepackage[cmex10]{amsmath}
\usepackage{amsfonts}
\usepackage{amssymb}
\usepackage{graphicx}
\usepackage{verbatim}
\usepackage{array}
\usepackage{multirow}
\usepackage{dcolumn}
\usepackage{xcolor}
\usepackage{url}
\usepackage{graphicx}
\DeclareGraphicsExtensions{.eps}
\usepackage{balance}

\usepackage[noadjust]{cite}
\usepackage{bbm}
\usepackage{float} 

\usepackage{mathrsfs}
\usepackage{color}
\usepackage{booktabs}
\usepackage{changes}

\newlength\myindent
\renewcommand{\algorithmicrequire}{\textbf{Input:}}
\renewcommand{\algorithmicensure}{\textbf{Output:}}

\newcommand\Tstrut{\rule{0pt}{2.6ex}}  
\begin{document}
%

\title{Linearized Bregman Iterations for Automatic Optical Fiber Fault Analysis}

%
\author{Michael~Lunglmayr~\IEEEmembership{Member,~IEEE,}~and~Gustavo~C.~Amaral
\thanks{}
\thanks{M.~Lunglmayr is with the Institute of Signal Processing, Johannes Kepler University, Linz, Austria (e-mail: michael.lunglmayr@jku.at).}
\thanks{G.~C.~Amaral is with the Center for Telecommunications Studies, Pontifical Catholic University of Rio de Janeiro, RJ, Brazil and with QuTech and Kavli Institute of Nanoscience, Delft University of Technology, Delft, The Netherlands (e-mail: gustavo@opto.cetuc.puc-rio.br).}
\thanks{Copyright (c) 2018 IEEE. Personal use of this material is permitted.  However, permission to use this material for any other purposes must be obtained from the IEEE by sending a request to pubs-permissions@ieee.org.}
}


\maketitle
\begin{abstract}
Supervision of the physical layer of optical networks is an extremely relevant subject. To detect fiber faults, single-ended solutions such as the Optical Time Domain Reflectometer (OTDR) allow for precise measurements of fault profiles. Combining the OTDR with a signal processing approach for high-dimensional sparse parameter estimation allows for automated and reliable results in reduced time. In this work, a measurement system composed of a Photon-Counting OTDR data acquisition unit and a processing unit based on a Linearized Bregman Iterations algorithm for automatic fault finding is proposed. An in-depth comparative study of the proposed algorithm's fault-finding prowess in the presence of noise is presented. Characteristics such as sensitivity, specificity, processing time, and complexity, are analysed in simulated environments. Real-life measurements that are conducted using the Photon-Counting OTDR subsystem for data acquisition and the Linearized Bregman-based processing unit for automated data analysis demonstrated accurate results. It is concluded that the proposed measurement system is particularly well suited to the task of fault finding. The natural characteristic of the algorithm fosters embedding the solution in digital hardware, allowing for reduced costs and processing time.
\end{abstract}

%

\IEEEpeerreviewmaketitle
\section{Introduction}
Optical fiber networks provide the backbone of today's information society. In the year 2014, already about 25 million kilometers of optical fiber had been installed worldwide, carrying over $80\%$ of the long distance data traffic \cite{Kumar2014fiber}. Even though  an excellent media for broad band communications, the mechanical vulnerability of the optical fiber might jeopardize its transmission capacity: fiber bending, fiber breaking, imperfect fiber splices, and corrupted connectors are examples of faults that compromise the power budget in a fiber optical link and, eventually, render the link inoperable. Therefore, a means of detecting, locating, and evaluating the magnitude of such fault events is unavoidable and of great interest to network operators. Consequently, many physical layer monitoring techniques have been developed over the years, with the Optical Time Domain Reflectometry (OTDR) \cite{BarnoskiAO1977} and the Optical Frequency Domain Reflectometry (OFDR) \cite{eickhoff1981optical} figuring as the most successful ones. 

The OTDR, which measures the Rayleigh backscattered power of a high-intensity optical probing pulse to determine a fiber's profile \cite{DericksonBOOK1998}, is typically used as the data acquisition system for fiber fault analysis. As the architectures and sizes of the optical networks change with the development of new technologies and increasing demand for higher data rates, the monitoring techniques must also evolve to remain compatible with the networks \cite{urban2018tutorial}. An example of the evolution of supervision techniques to meet requirements of new architectures is the wavelength-tunable OTDR, which allows for the supervision of wavelength division multiplexing (WDM) optical networks \cite{amaral2014wdm}. An example monitoring system for multiple fibers using an OTDR device is described in \cite{HannMonitor2006}.

In this context, a natural evolution is the development of mathematical tools capable of automatically identifying the physical layer faults from the raw supervision data. This can eliminate the necessity of human resource allocation to this exhaustive process, which, in turn, reflects into minimized expenses and reduced downtime in the network caused by the scheduling of in-field repair units \cite{IEC2014}. Apart from this operational advantage related to optical network management, location and estimation of losses in an OTDR profile can also be used for sensing applications: in \cite{WuBend2016}, a system for measuring the mechanical displacement of a fiber by measuring the losses due to fiber bends has been presented; in \cite{ScullyPH2007}, fiber losses sensed by a Photon-Counting OTDR are used to measure the pH value of a liquid. Naturally, a methodology able to locate and characterize faults in a fiber is general enough so that it can be adapted to either application.

The previously mentioned fiber faults, or any other fault in a fiber that causes the transmitted power to drop in an individual location, have been shown to manifest as trend breaks, or steps, in the digitized OTDR measurement data \cite{AmaralJLT2015,WeidJLT2016}. Therefore, the location of a step can be directly associated to the location of the corresponding fiber fault. In other words, detecting faults in an OTDR profile is equivalent to detecting steps in the trend of the digitized data. If an algorithm can be responsible for such identification, automatic location of potential faults could be performed.

\begin{figure}[htbp]
\center
\includegraphics[width=0.9\linewidth]{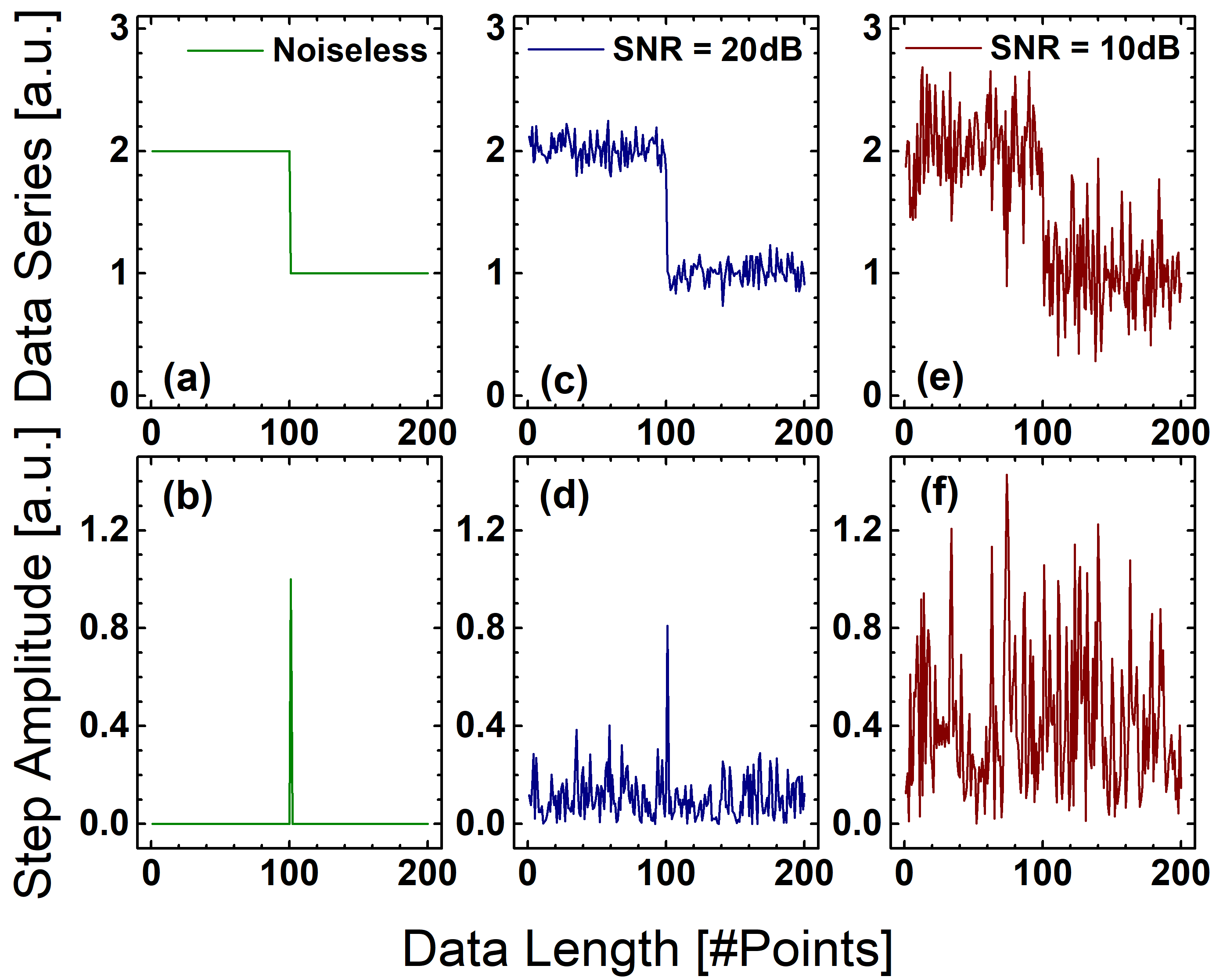}		
\caption{Trend break detection using a discrete derivative and peak detection method. The result is compared under three signal-to-noise ratio environments: infinite; 20dB; and 10dB. The upper panels show the original data and the lower panels show the results, where a peak identifies the trend break.}
\label{fig:motiv}	
\end{figure}

The reliable identification of steps in a data series is a matter that attracts substantial interest in areas such as: detection of disease outbreaks \cite{wagner2001emerging}; analysis of geophysical signals \cite{basseville1983desgin}; surveillance \cite{clavel2005events}; speech processing \cite{rabiner1989tutorial}; macroeconomics \cite{karney1983long}; fraud detection \cite{bolton2002statistical}; or even in biological machines \cite{Carter2008machina}. Although this problem is extremely simple in a noiseless scenario, it becomes quite challenging in the presence of noise. In Fig. \ref{fig:motiv}, this feature is exemplified when a trend break is present under different signal-to-noise ratio (SNR) conditions. While in the first case (Fig. \ref{fig:motiv}-a and -b) taking the discrete derivative of the original data and searching for a peak in the outcome is sufficient to precisely identify the fault, as the SNR decreases (Fig. \ref{fig:motiv}-c, -d -e, and -f), the process yields poor results. 

For this reason, and due to its relevance, signal processing techniques have been developed \cite{brown1975techniques, rea2010identification, storath2014jump, candes2008enhancing, kim2009ell_1} to attack the problem of multiple, unknown, trend break detection in the presence of noise. Specifically in OTDR data acquisition, noise is present due to the power losses along the fiber that causes the measurement SNR to decrease and to the intrinsic scattering noise known as Coherent Rayleigh Noise (CRN) \cite{DericksonBOOK1998}. Also specific to the context of fiber monitoring is the fact that, from the point of view of a network operator, being able to identify and repair a fault before a user complaint is registered is extremely advantageous, the technique must be highly sensitive to the presence of a fault. On the other hand, triggering a repair unit to repair a fault that does not exist is economically harmful, so the technique must also exhibit high specificity in fault detection. 

In a recent work \cite{WeidJLT2016}, the \textit{Adaptive $\ell 1$ Filter}, an algorithm based on sparse estimation, was shown to outperform other techniques with respect to the aforementioned expected characteristics. Despite being an excellent method for trend break detection under noise, the Adaptive $\ell_1$ Filter reveals some shortcomings with respect to an implementation in an embedded system for automated fiber measurements. It relies on a coordinate descent algorithm, with a greedy candidate selection process similar to the algorithm described in \cite{Friedman}. Even though the simplicity of the model used for step detection enables low complexity estimation algorithms, the pre-processing step required for the Adaptive $\ell 1$ Filter leads to multiplication-heavy algorithms, increasing its complexity. In addition, as the authors of \cite{Friedman} point out, to ensure stability, the algorithm requires calculating complete solutions paths with different levels of sparsity, which may compromise its processing time. On a different proposal \cite{OTDRTIMI, OTDRTIMII}, detection of splice faults in an OTDR profile has been demonstrated with a digital signal processing approach based on a Gabor series representation and rank-1 matched subspace detection. Although the excellent detection capability demonstrated, the computational complexity of this approach is cubic in the number of measurement samples thus also limiting its implementation in an embedded measurement system for long fibers.

In the context of sparse estimation, Linearized Bregman Iterations (LBI) offer a low complexity and high precision approach for solving the combined $\ell 1/\ell 2$ minimization problem \cite{LBOsher, Osher2, Osher4, Yin, Yinlin, lunglmayr2016efficient}. By adapting the LBI algorithm to take into account the characteristics of the trend break detection problem, a signal processing methodology was developed exhibiting better timing and accuracy when compared to state-of-the-art solutions, namely the \textit{Adaptive $\ell_1$ Filter}. In this paper, the Linearized Bregman Iterations framework is presented as a candidate for an automatic trend break detection technique. Through an in-depth analysis focused on the particular case of fault detection in noisy OTDR profiles, figures of merit such as sensitivity, specificity, and processing time have been drawn in simulated and real-life environments. For comparison analysis between the proposed algorithm and the Adaptive $\ell_1$ Filter, both algorithms have been implemented in the Julia language. The conclusion is that the LBI algorithm is a promising candidate not only for this particular problem, but for the more general issue of trend break detection in noisy data. Furthermore, since the prospect of automated operation promotes the urge of reduced complexity so that a low-cost embedded processing unit such as an FPGA can be responsible for real-time monitoring of the fiber links, this point has also been investigated and the proposed algorithm is shown to offer promising characteristics for FPGA embedding.

The paper introduces the basic concepts of OTDR operation and the mathematical model of the Linearized Bregman Iterations algorithm in Section II. Section III is responsible for highlighting the modifications on the original Linearized Bregman Iterations algorithm necessary to adapt it to the fault finding problem and thus creating a consistent parameter-free processing algorithm. With the background developed in the previous sections, a testbench experimental result is tested with the proposed algorithm. The results allow the development of a framework of extensive simulated profiles and the results both from real fiber measurements and simulated fiber profiles is detailed in Section IV. Finally, Section V concludes the work summarizing the findings as well as laying the path for future developments.

\section{OTDR and LBI - Basic Concepts}
\label{sect:2}
\begin{figure}[htbp]
\center
\includegraphics[width=0.9\linewidth]{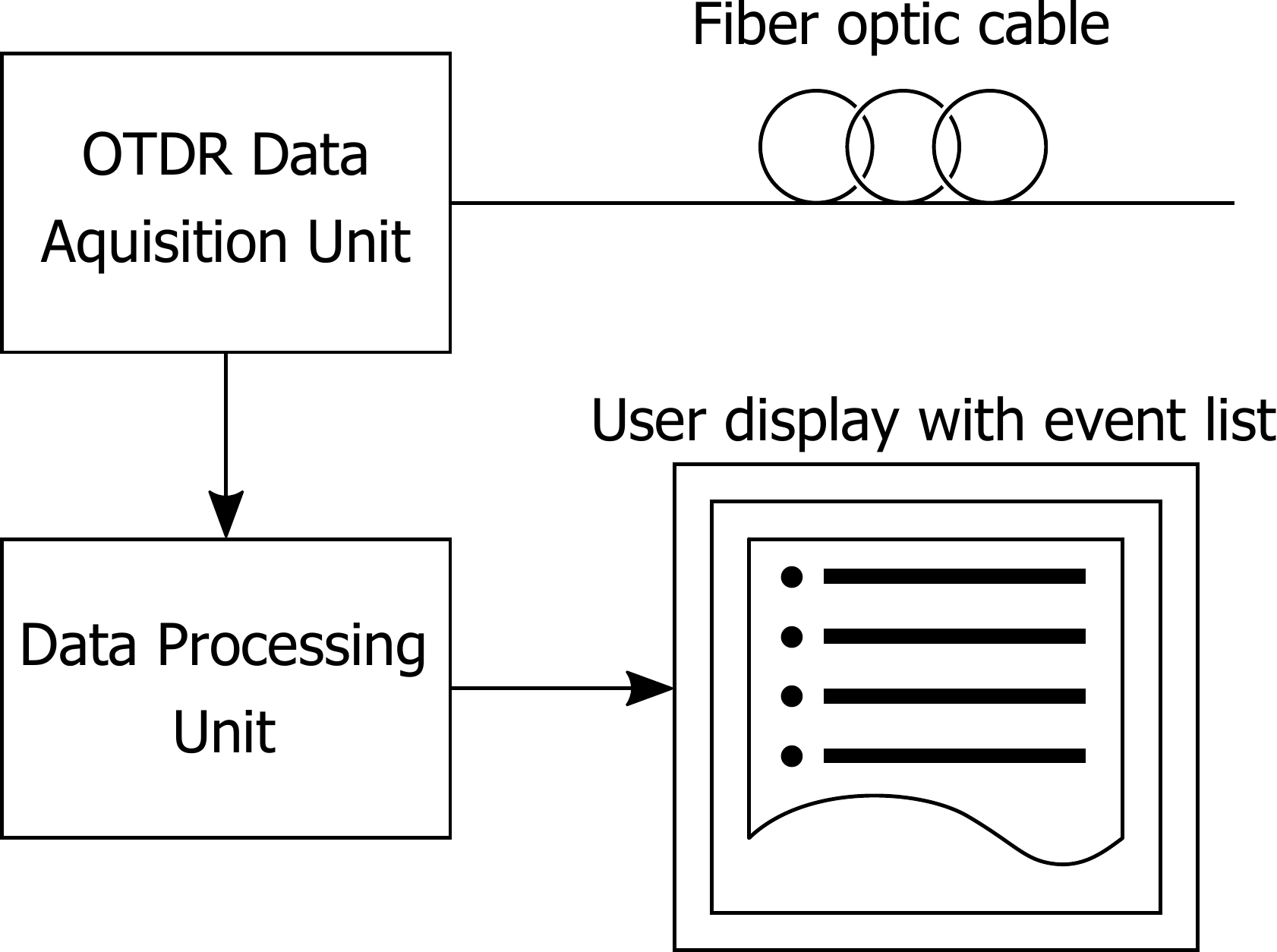}		
\caption{Block diagram of the proposed measurement system with its desired output: an ``event list" containing positions and magnitudes of events along the optical fiber link. The data acquisition subsystem represents the Photon-Counting OTDR of \cite{AmaralJLT2015}.}
\label{fig:measSystem}	
\end{figure}

\begin{figure}[htbp]
\center
\includegraphics[width=0.9\linewidth]{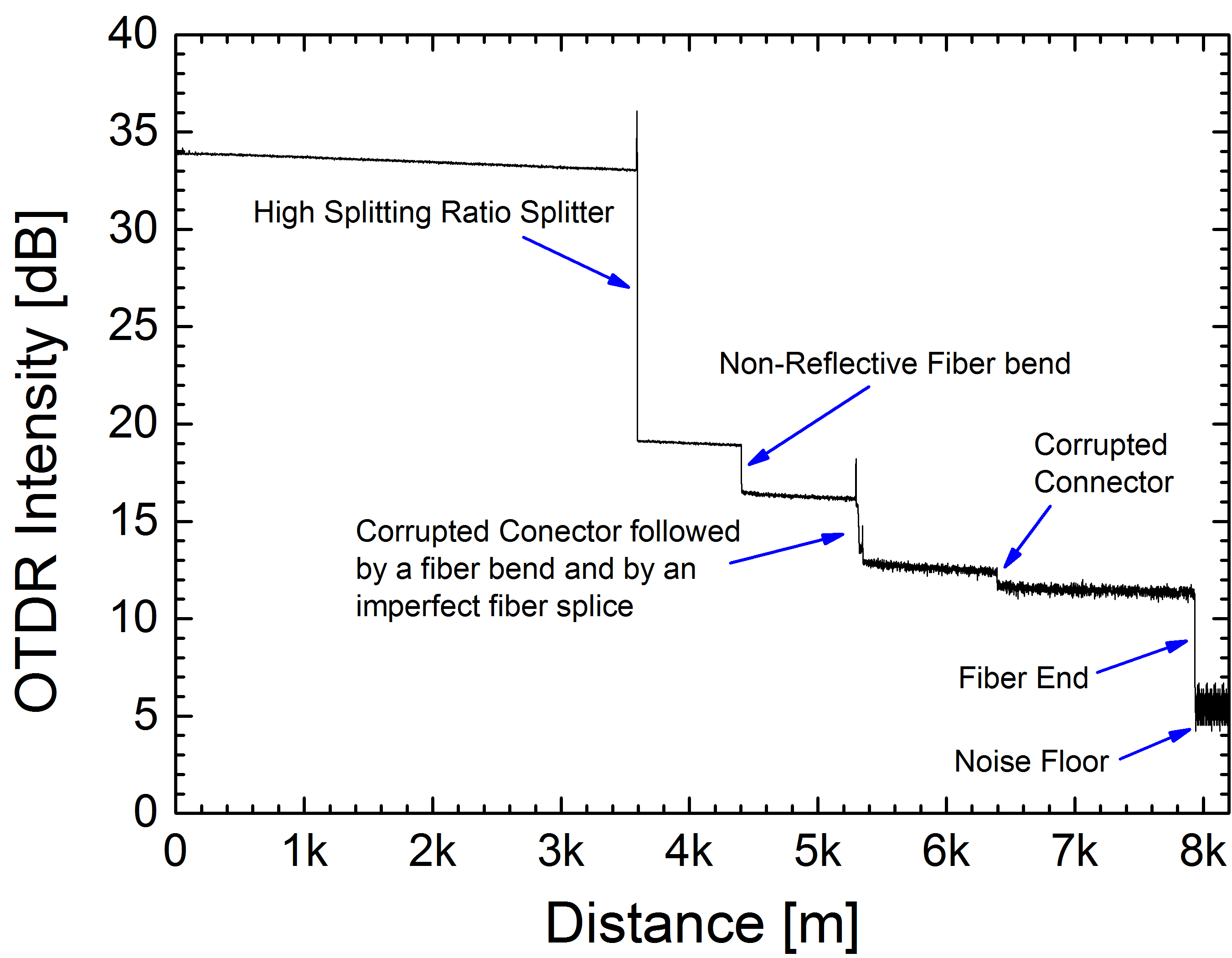}		
\caption{Example of OTDR fiber profile highlighting the most common fiber fault causes. The profile was acquired with a tunable Photon-Counting OTDR setup, presented in detail in \cite{AmaralJLT2015}.}
\label{fig:exampleOTDRTrace}	
\end{figure}

Optical Time Domain Reflectometry is a single-ended reflectometry measurement technique that consists of measuring the Rayleigh backscattered portion of a light pulse traversing the fiber as a function of time. The OTDR provides the fiber profile by associating the measured data to the pulse's round-trip time, which, combined with the knowledge of the fiber's refractive index and the speed of light, can be translated into distance. The fiber profile is, thus, a descending line in logarithmic scale with a slope equal to the fiber's attenuation coefficient \cite{agrawal2002book, BarnoskiAO1977, DericksonBOOK1998}. Any event which causes optical power loss is interpreted accordingly, i.e., the  aforementioned effects of fiber bending, fiber breaking, imperfect fiber splices, and corrupted connectors can be identified by an abrupt level shift in the signal profile. The goal would be to identify such level shifts and associate their positions and magnitudes to faults. The proposed automated measurement system that provides such results is depicted schematically in Fig. \ref{fig:measSystem}, where the Photon-Counting OTDR presented in \cite{AmaralJLT2015} is used as the data acquisition subsystem. In Fig. \ref{fig:exampleOTDRTrace}, an example OTDR fiber profile acquired with this system is presented, indicating the most common effects associated to power loss and/or fiber faults. After the data acquisition step, the Linearized Bregman-based processing unit analyses the resulting OTDR profile and provides the event list to the user. 

Given the OTDR fiber profile such as the one in Fig. \ref{fig:exampleOTDRTrace}, eventual power losses that could possibly compromise the link's transmission capacity will be identified by trend breaks. Although some of the power losses are expected -- for instance the splitters and connectors necessary for the network architecture -- some are spurious or unexpected and often arise due to the mechanical fragility of the optical fiber \cite{DericksonBOOK1998}. The power balance of the optical link, which takes into account the input optical power and the detector margin for error-free reception, may be able to cope with some of these losses but, eventually, they might render the link inoperable. The task, therefore, is to rapidly detect the presence of such breaks with the highest accuracy possible (within the probing method's limitation). This way, scheduling of in-field repair units can be accelerated and managing costs of the network supervisor can be reduced as well as any down-times experienced by the network user. 

It should be noted that the spatial resolution employed by the measuring technique will limit the achievable accuracy of an algorithm that processes the output digitized data. The OTDR, or any other conceivable fiber monitoring apparatus, albeit offering an excellent tool for fiber monitoring, poses a trade-off in terms of its measurable distance and resolution: in order to investigate distant parts of the fiber, the probe pulse is enlarged -- so to carry more optical power -- and the spatial resolution is thus reduced. If a higher resolution is required, on the other hand, the probe pulse width is reduced at the expense of reduced reach. For that reason, a signal processing approach will not be able to locate a fault with a precision higher than the physical parameters of the measuring apparatus allow for. Accuracy for automatic fault detection should be interpreted as ``as accurate as the physically available spatial resolution". The signal processing approach deals with points in a data series and it is left to the measurement apparatus to link the digitized data points to a physical position or distance.

The concept behind the automatic identification of fault positions is to interpret the OTDR profile as a weighted sum of step functions and then find the steps and respective weights that best explain the original data series. The profile depicted in Fig. \ref{fig:exampleOTDRTrace} clarifies this point, as, after each highlighted event, there is a sharp decrease in the measured power. Step functions are uniquely defined by their amplitude and the position beyond which they assume a non-zero value. This allows to cast an arbitrary profile, under this interpretation, in the following form:
\begin{equation}
{y(z)} = \alpha \sum_{i \in F}a_i {u}\left(z-z_i\right),
\label{eq:linComb}
\end{equation}
where ${y(z)}$ is the ideal (noise-free) fiber profile, $z$ denotes the distance inside the fiber, $F$ is the set of fault indices, $a_i$ are the weights (or magnitudes) of the faults, and $z_i$ its positions. The step function is represented by the Heaviside function $u\left(z\right)$ and the factor $\alpha$ stands for the linear slope of the fiber attenuation. For a measurement application,
$y(z)$ is typically sampled equidistantly and the samples are combined into a vector $\bf y$.

Identification of the set $F$ of fault indices is the goal of the fault finding problem, which can be assumed to be sparse in this particular application. 
Fig. 2 is an example of the sparse characteristic of the problem since the total number of possible fault positions is 8000 (each entry in the dataset, or the spatial resolution of the acquisition system, corresponds to one meter), but the size of the set $F$ is $7$. Indeed, from a practical point of view, it makes sense that the number of events should be much smaller than the points in the data series: a high number of fault events translates into high optical losses and, as commented above, the inoperability of the optical link. 
A remarkable point that arises from this reasoning is that, given the spatial resolution, one has access to all possible positions in the digitized data series where a trend break might appear. Thus, a naive approach would be to check all the possible combinations of step functions until the best approximation to the measured signal is found following (\ref{eq:linComb}). Such a procedure, however, is known to belong to the class of the so-called NP-hard problems \cite{natarajan1995sparse} and, as such, is computationally intractable. A relaxation of this extremely difficult problem is to use the $\ell_1$ norm instead, which allows one to employ algorithms with manageable complexity that output sparse results for a broad range of practical problems \cite{candes2006robust}.

To estimate the fault positions within a given OTDR profile, a dictionary of candidate vectors is provided, where the name dictionary is used to emphasize that candidate vectors do not have to be linearly independent \cite{candes2008dict}. There can be, as it is used in this work, even more candidate vectors than measurement values, in which case it is often referred to as an overcomplete dictionary \cite{candes2008dict}. The $\ell_1$ minimization problem where the dictionary of candidates is provided is known as the basis pursuit problem which, in its matrix form, can be represented as:
\begin{equation}
\min_{\boldsymbol{\beta}} ||\boldsymbol{\beta}||_1 \hspace{0.2cm} s.t. \hspace{0.2cm} {\bf A}\boldsymbol{\beta} = {\bf y},
\label{eq:basisPursuit}
\end{equation}
where $\bf A$ is the dictionary, with each column of $\bf A$ being a dictionary or candidate vector. $\boldsymbol{\beta}$ are the coefficients of the dictionary vectors, which represent the amount of contribution, or weight, of a candidate to the fiber profile data series ${\bf y}$.

Based on (\ref{eq:linComb}), (sampled) step functions shifted at all possible sample points will be included as candidate vectors in matrix $\bf A$. In addition, the estimate of the attenuation coefficient is taken into account by also including a single linear candidate with unitary slope into the dictionary ($\bf A$ is described in more detail in Sect.~\ref{sect:3}). Ultimately, estimation of the attenuation as well as the level shifts is the estimation of the coefficient vector $\boldsymbol{\beta}$ where the non-zero coefficients identify the step locations and, thus, the position of the faults in the profile. 

Linearized Bregman Iterations (LBI) form a class of implementation-efficient algorithms that solve the combined $\ell_1/\ell_2$ problem of the form:
\begin{equation}
\min_{\boldsymbol{\beta}} \lambda ||\boldsymbol{\beta}||_1 + \frac{1}{2\alpha}||\boldsymbol{\beta}||_2^2 \hspace{0.2cm} s.t. \hspace{0.2cm} {\bf A} \boldsymbol{\beta}={\bf y}.
\label{eq:linearizedBregman}
\end{equation}
As e.g. shown in \cite{Augmented, LBOsher, LorenzNoise}, Linearized Bregman- based algorithms also give meaningful results in the presence of noise, i.e., even in cases when ${\bf A}\boldsymbol{\beta}={\bf y}$ does not have a solution. In 
such scenarios, it can be shown that the squared error norm of an estimate $\boldsymbol{\beta}^{(k)}$ at iteration $k$ (details on the algorithm are given below) is reduced by following iterations as long as $\|{\bf A}\boldsymbol{\beta}^{(k)}-{\bf y}\|_2$ is higher than the noise level \cite{Augmented, LBOsher}. An interesting observation is that, adding the $\ell_2$ norm in the cost function above, leads to very low complexity solution algorithms \cite{LBOsher, Yinlin} that can be implemented very efficiently in digital hardware \cite{ISCAS2017}. Furthermore, it can be noted that, with the addition of the $\ell_2$ norm term, a correct balancing of $\lambda$ and $\alpha$ allows for the solution of the basis pursuit problem (\ref{eq:basisPursuit}) within the framework of the LBI (\ref{eq:linearizedBregman}) in an efficient manner. For simplicity, we set $\alpha = 1$, leaving only the parameter $\lambda$ to adjust the weight of the $\ell_1$ versus the $\ell_2$ norm. 

Despite their low complexity, Linearized Bregman Iterations still require matrix vector multiplications. A further simplification, however, called Sparse Kaczmarz, only requires vector operations while still maintaining convergence \cite{lorenz2014sparse}. The basic algorithm of Sparse Kaczmarz is shown in Algorithm \ref{alg:Kaczmarz}, where the Kaczmarz simplification has been used: instead of the full matrix $\bf A$, only a single row of $\bf A$ is used in a single iteration of the algorithm. Although this algorithm represents an adaptation of the original Linearized Bregman Iterations algorithm described in \cite{LBOsher}, in the following, references to the LBI algorithm should be interpreted as to the Kaczmarz variant. This will also prevent confusion between LBI-based Sparse Kaczmarz algorithms and other algorithms also called Sparse Kaczmarz, e.g., as described in \cite{TheOTHERSPARSEKACZ}.

\begin{algorithm}[ht]
\caption{Linearized Bregman-based Sparse Kaczmarz}
\label{alg:Kaczmarz} 
\small
\begin{algorithmic}[1]
\State{$\boldsymbol{\beta}^{\left(0\right)} \leftarrow {\bf 0} $}
\State{${\bf v}^{\left(0\right)} \leftarrow {\bf 0}$}
\For{$k=1..N$}
\For{$j=1..p$}
\State{$\boldsymbol{\beta}_j^{\left(k\right)}\leftarrow $ shrink $\left(v_j^{\left(k\right)},\lambda\right)$}
\EndFor
\State $i \leftarrow ( (k - 1) \text{ mod }p ) + 1$ \Comment {cyclic re-use of rows of ${\bf A}$}
\State{${\bf v}^{\left(k+1\right)}\leftarrow {\bf v}^{\left(k\right)}+\frac{1}{\|{\bf a_i}\|_2^2}{\bf a_i}\left(y_i-{\bf a_i}^T\boldsymbol{\beta}^{\left(k\right)}\right)$}
\EndFor
\end{algorithmic}
\end{algorithm}
%
The LBI-based Sparse Kaczmarz algorithm can be seen as a steepest descent approach using an approximate gradient derived from a single row of ${\bf A}$ and a single measurement value (line $8$) combined with a \textit{sparsifying} operation (line $5$), while the term $\frac{1}{\|{\bf a_i}\|_2^2}$ can be seen as the \textit{step width} of the descent.  The sparsifying operation is due to the definition of the shrink function: shrink$(v, \lambda) = \text{max}(|v|-\lambda,0) \cdot \text{sign}(v)$, that sets values of $v$ with a magnitude smaller than $\lambda$ to zero. 

In general, a sensible relation when using the Linearized Bregman Iterations is, as commented above, the balance between the $\ell_1$ and $\ell_2$ norm translated by $\lambda$. When adapting the Sparse Kaczmarz algorithm to the OTDR fault detection problem, this parameter turned out to have a much lower sensibility than it is observed for general $\ell_1/\ell_2$ estimation problems, as will be discussed in the following Section.

\section{Analysing OTDR Fiber Profiles with the Linearized Bregman Iterations}
\label{sect:3}
The first, most crucial aspect of adapting the Bregman methodology to any basis pursuit problem, is the definition of the matrix $\bf A$, the dictionary of candidates that will compose the signal of interest as a linear combination. In the specific case of OTDR profiles, the candidates are step functions, which correspond to discontinuous descends of the profile magnitude. As described in Sect.~\ref{sect:2}, due to an intrinsic attenuation of light propagation through optical fibers, the matrix $\bf A$ must also account for a negative slope. The final form of ${\bf A}$ is:
\begin{equation}
{\bf A} = \begin{bmatrix}
 1         & 1        & 0         & 0      & \cdots &  0        & 0 \\
 2         & 1        & 1         & 0      & \cdots &  0        & 0 \\
 3         & 1        & 1         & 1      & \cdots &  0        & 0 \\
 \vdots & \vdots& \vdots &         & \ddots & \vdots & \vdots \\
 p-2     & 1        & 1         & 1      & \cdots & 1         & 0 \\
 p-1         & 1        & 1         & 1      &  \cdots& 1         & 1
\end{bmatrix},\label{eq:matrixX}
\end{equation}
with $n=p-1$ rows. The structure of ${\bf A}$ allows for a significant simplification that can be incorporated into the Sparse Kaczmarz algorithm, preventing the need to store and load this matrix, and is as follows.

When analysing a row ${\bf a}_i^T$ of ${\bf A}$, one can see that this row consists of the number $i$ as its first element followed by $i$ one elements and, finally, by $p-i-1$ zero elements. This allows to formulate the inner product calculation of ${\bf a}_i^T{\boldsymbol{\beta}}$ efficiently as
\begin{equation}
{\bf a}_i^T{\boldsymbol{\beta}} = i \beta_1 + \sum_{s=2}^{i+1} \beta_s, 
\end{equation}
as well as the calculation of the squared norm $\|{\bf a}_i\|_2^2$ as
\begin{equation}
\|{\bf a}_i\|_2^2 = i^2 + i.
\end{equation}
Considering that this norm is used as step width in the LBI algorithm ($\frac{1}{\|{\bf a}_i\|_2^2} = \frac{1}{i^2 + i}$), the squared part
$ i^2$, originating from the first column, contributes most significantly to the norm. This means that the contribution of the first column scales down the step width, limiting the algorithm's convergence speed. For this reason, a scaling factor $\sigma$ for the first column of ${\bf A}$ is used in
\begin{equation}
{\bf a}_i^T{\boldsymbol{\beta}} = \sigma i \beta_1 + \sum_{s=2}^{i-1} \beta_s, 
\end{equation}
as well as in
\begin{equation}
\|{\bf a}_i\|_2^2 =  \sigma^2 i^2 + i,
\end{equation}
leading to larger step widths and faster convergence. One point to consider is that, with this scaling factor, the first element of the output of the algorithm is also scaled by $\sigma$. Since the algorithm counteracts this scaling and implicitly multiplies by $1/\sigma$ during the estimation process, the first element has to be multiplied by $\sigma$ at the end of the algorithm for compensation. Considering a future hardware implementation, the value of $\sigma$ was selected as a power of two, which allows for replacing the multiplication by a shift operation. Furthermore, for large values of the step width $\frac{1}{\|{\bf a}_i\|_2^2}$, the value of $\sigma$ should be chosen as small as possible. However, a compromise between large step widths and the required numerical dynamic range, especially when considering a fixed point implementation, has to be found. For this reason $\sigma = 2^{-10}= 1/1024$ was selected allowing for a fast convergence while still keeping the required dynamic range manageable in a future hardware implementation.

These considerations allowed for the development of the following Algorithm 2. In order to clarify the steps of the algorithm, a flowchart accompanies the pseudo-code's main steps. This flowchart shows the simple iterative nature of the algorithm, consisting mainly of a successive application of the described Sparse Kaczmarz iteration: an approximate gradient step followed by the evaluation of the shrink function. The structure displayed in Algorithm 2 fosters a low complexity implementation (complexities of Algorithm 2 as well as of the Adaptive $\ell_1$ Filter \cite{WeidJLT2016} are discussed and compared in Sect.~\ref{sect:cmplx}).\\
\begin{center}
\includegraphics[width=0.9\linewidth]{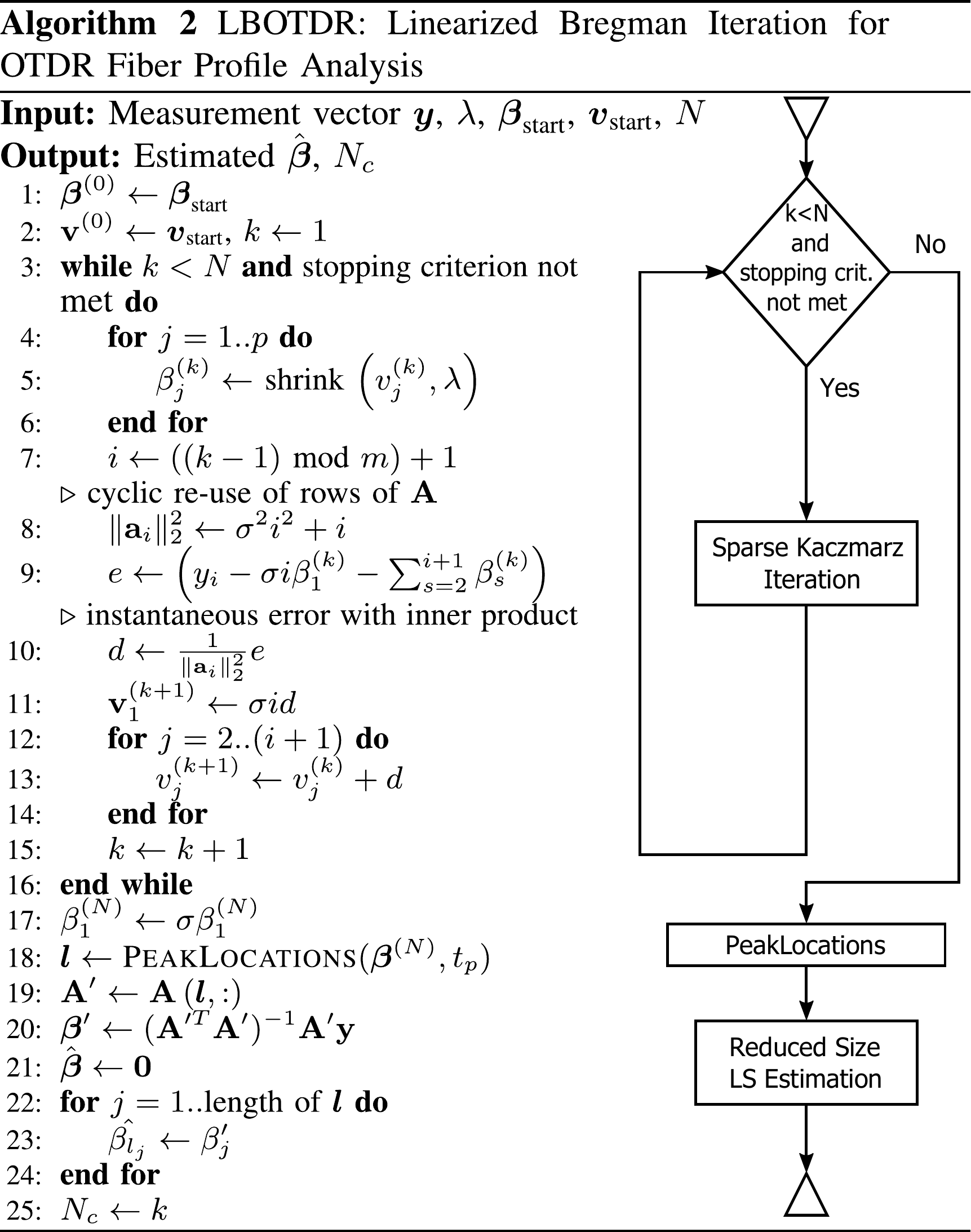}
\end{center}
\addtocounter{algorithm}{1}

Due to the structure of matrix ${\bf A}$, the operations that require access to its rows can be directly encoded in the algorithm without the necessity to load $\bf A$, allowing Algorithm 2 to perform the LBI with reduced complexity. The function $\textsc{PeakLocations}$, as shown in Algorithm \ref{alg:betapeak}, returns a vector containing those positions where the sign of the difference between consecutive positions changes, which is used to detect local maxima in the estimated $\boldsymbol{\beta}$. The detection of peaks works, in Algorithm 2, as an efficiency booster since it lessens the sensitivity of the algorithms to the parameter $\lambda$. In addition, as Linearized Bregman-based algorithms show the tendency to require more iterations when using larger $\lambda$ values \cite{LorenzNoise, lunglmayr2016efficient}, this allows for fast convergence while still maintaining sparse estimation results. After detecting the peaks, their respective positions are used to select the corresponding columns of $\bf A$ to form matrix ${\bf A}'$. Then, a least squares estimation is performed in the reduced subspace (spanned by the peak positions) of the larger column space of $\bf A$. We describe the stopping criterion of Algorithm 2 in the next subsection.

\begin{algorithm}[ht]
\caption{PeakLocations}
\label{alg:betapeak}
\algorithmicrequire{$\boldsymbol{\beta}$, $t_p$}\\
\algorithmicensure{Vector of peak locations $\boldsymbol{l}$}\\
\vspace{-.5cm}
\begin{algorithmic}[1]
\small
\State $S \gets \emptyset$
\For{$j=2..(p-1)$}
\If {$\text{sign}(\beta(j+1)-\beta(j)) \neq \text{sign}(\beta(j)-\beta(j-1))$}
  \State $S \gets S \cup \{j\}$	
\EndIf
\EndFor
\State Remove all indices $i$ from $S$ where $\beta_i < t_p$
\State $S \cup \{1\}$ \Comment{always add slope coefficient}
\State form vector  $\boldsymbol{l}$ of all elements of $S$ in ascending order
\end{algorithmic}
\end{algorithm}

In $\textsc{PeakLocations}$, the first element of the estimated $\boldsymbol{\beta}$ is always included in the output ensuring that the analysis always considers the fiber profile's slope represented by the first column of $\bf A$. This phenomenological observation corresponds to the fact that a non-zero slope must always be present since no fiber exhibits zero attenuation. Furthermore, in the implementation with the Julia language, the least squares estimation of line 20 in Algorithm 2 is performed via a matrix inverse. In a hardware implementation, the same operation can be efficiently performed via Kaczmarz-like $\ell_2$ estimation algorithms \cite{ALS}. Since the algorithm used for results comparison also requires such a least squares estimation step, this Kaczmarz-like $\ell_2$ estimation has not been included for comparison's sake. It should be noted, however, that the LBI algorithm provides the advantage that existing hardware elements can be easily re-used due to the fact that, by setting $\lambda$ to zero, the Sparse Kaczmarz algorithm becomes the ordinary Kaczmarz algorithm.

An important aspect of an algorithm directed to the problem of automatic multiple trend break detection is the convergence criteria: the algorithms' processor must reach a state at which the results are considered to be trustworthy and further processing would only minorly change the estimation result while still consuming processing resources. Inspecting the pseudo code presented in Algorithm \ref{alg:Kaczmarz}, one can see that the Sparse Kaczmarz can be implemented without any convergence criterion other than the selection of the number of iterations $N$ that shall be ran by the algorithm. Even though, for some applications, an \textit{a priori} determined $N$ suffices for a convergence criteria, it is often helpful to add a convergence criterion based on the quality of the estimate, which allows reduced number of iterations until convergence and guarantees a robust and consistent output.

\subsection{Stopping Criterion}

\begin{figure}[htbp]
\center
\includegraphics[width=0.9\linewidth]{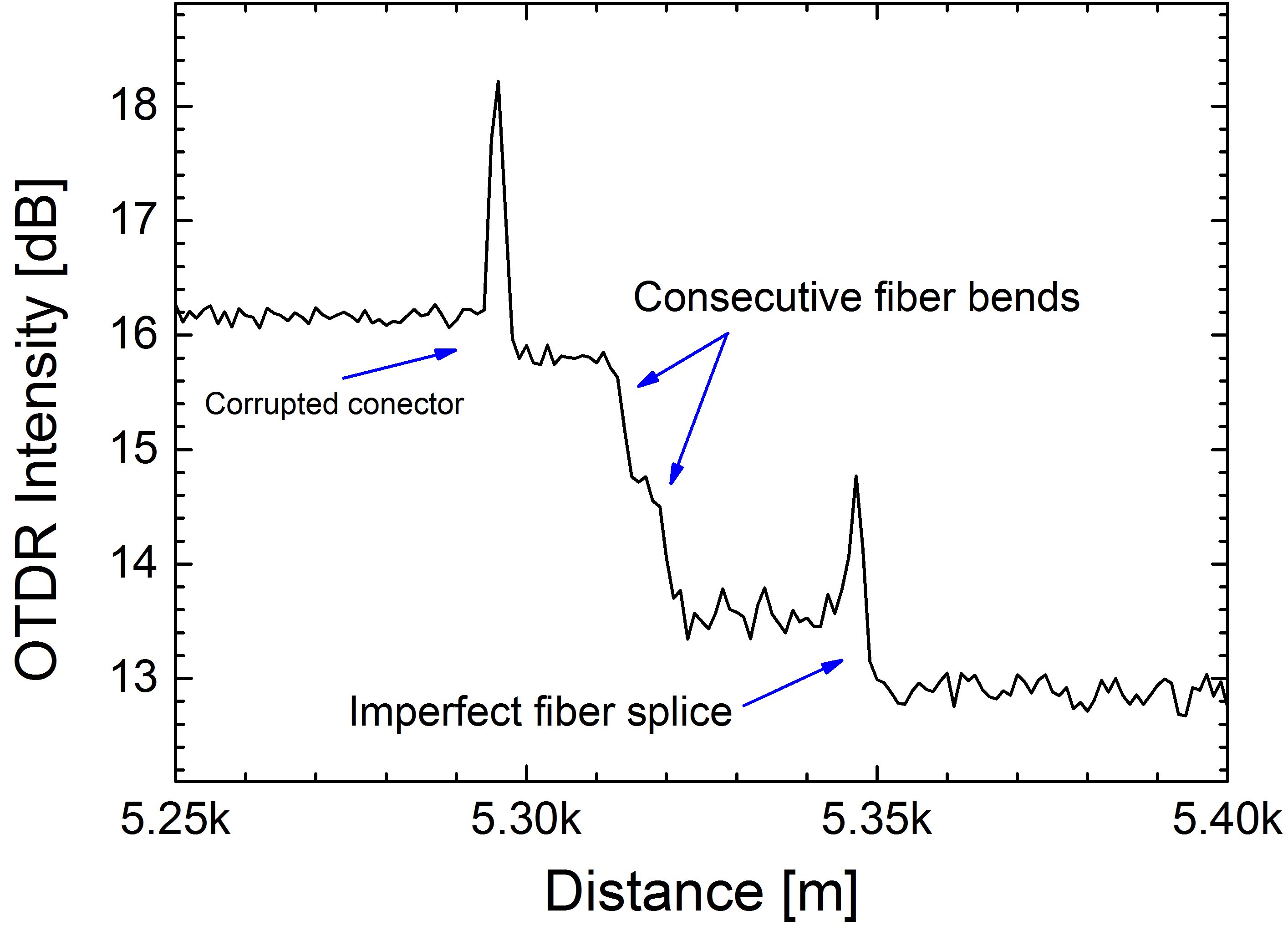}		
\caption{Magnified view of an OTDR profile where multiple faults can be distinguished within very close spacing from each other. Such events are hard to be detected, and a fixed number of iterations might create a pitfall in this case.}
\label{fig:zoomOTDRTrace}	
\end{figure}

In Fig. \ref{fig:zoomOTDRTrace}, a magnified view of the OTDR profile of Fig. \ref{fig:exampleOTDRTrace}, focusing on the region between 5.25 km and 5.4 km, is presented. Within this region, several faults of different magnitudes can be found very close to each other, almost to the limit of the measurement apparatus's spatial resolution. This particular set of faults is hard to be detected by an analysis algorithm. To ensure high quality estimation results such exceptional cases would have to be considered when setting the maximum number of iterations $N$. For scenarios that do not exhibit such an exceptional fault configuration, on the other hand, running the algorithm for the same number of iterations might lead to an unnecessary expense of processing time. Needless to say, this information is not available \textit{a priori} to the algorithm, so choosing $N$ based on the characteristics of the profile is not possible. For that reason, a simple but effective convergence criteria is introduced as an addition to the original Linearized Bregman Iterations by resorting to a practical implementation factor.

In practice, faults smaller than a given value -- usually expressed in dB to follow the OTDR logarithmic intensity representation -- are negligible, i.e., their presence, even though potentially indicating a process that can lead to a jeopardizing fault, is not relevant enough to trigger a repairing unit. In addition, due to the presence of noise, it becomes extremely challenging to detect trend breaks with a magnitude comparable to the noise level. This simple concept may be used to introduce a \textit{minimum detectable loss} for the analysis algorithm: the minimum loss the algorithm should be able to detect, on average, over all the sampled data points. Throughout the result analysis, the minimum detectable loss was arbitrarily set to $0.125$ dB, an extremely strict value that, depending on the user's will, can be raised for faster convergence. It is important to note that setting a value for the minimum detectable loss does not guarantee that the algorithm will find all trend breaks with magnitudes higher than its value but, rather, provides a compromise between the algorithm's performance and its elapsed time.

\subsection{Using multiple $\lambda$ values}

Linearized Bregman-based algorithms use the parameter $\lambda$ in the cost function to trade sparsity (considered by the $\ell_1$ norm) for model fitting error (considered by the squared $\ell_2$ norm part). In general, a selection of $\lambda$ is application-specific and no closed form solutions for choosing its optimal value exist. Even though smaller values of $\lambda$ lead to faster convergence, it comes at the expense of lower sparsity in the results. In order to combine these two apparently contradicting characteristics, a set of $\lambda$ values together with an approach called hot-starting \cite{CAMSAP} has been implemented.

Algorithm 2 starts with an empirically determined value of $\lambda=0.5$ and runs until convergence is reached. Using $\lambda=0.5$ is also beneficial for a fractional precision fixed point implementation of the LBI algorithm, as has been described in \cite{ISCAS2017}. As previously commented, this pre-set value of $\lambda$ yields high-quality results due to the $\textsc{PeakLocations}$ function, which will be demonstrated in Sect. \ref{sect:results}. A new value of $\lambda$ is then selected from a set $S_\lambda$ and Algorithm 2 runs again with hot-started vectors $\boldsymbol{\beta}$ and $\boldsymbol{v}$. $S_\lambda$ is constructed as a grid from the initial value of $\lambda=0.5$ up to $\lambda_{max}$ -- a $\lambda$ value that would already yield a trivial (all-zero) solution of (\ref{eq:linearizedBregman}) \cite{Friedman, lambdamax}. The hot-starting approach allows re-using the already calculated result of the LBI in a consecutive run with a higher lambda so that a significantly small number of iterations may still lead to convergence. Even though the vector $\boldsymbol{\beta}$ of a previous run can be directly used in the consecutive run, vector $\boldsymbol{v}$ must be adjusted to the new $\lambda$ value. This can be performed through \cite{CAMSAP}:
\begin{align}
v_j = (|\beta_j| + \lambda) \text{sign} (\beta_j),
\end{align}
for all $j = 1, \ldots, m$. In order to constrain the processing time taken for the evaluation of higher $\lambda$ values, a maximum number of iterations $N=0.1N_c$ is allowed for each subsequent run after the first, where $N_c$ is the total amount of iterations of the first run that led to convergence.

Selecting the model that best explains the original signal while maintaining the expected sparsity is performed by the so-called Bayesian Information Criterion (BIC) \cite{schwarz1978BIC}:
\begin{align}
\text{BIC} (\boldsymbol{y}, \hat{\boldsymbol{\beta}}) = \|\hat{\boldsymbol{\beta}}\|_0 \; \text{log}(p) + p \; \text{log}( \|\boldsymbol{y} - \boldsymbol{A} \boldsymbol{\beta}  \|_2^2 /p ).
\end{align}
The $\boldsymbol{\beta}$ vector with the smallest BIC value is output as $\hat{\boldsymbol{\beta}}_\text{best}$ in Algorithm \ref{alg:LBlambda}. In the following Section, the performance achieved by Algorithm \ref{alg:LBlambda} is presented and analysed.

\begin{algorithm}[ht]
\caption{ Model Selection with different $\lambda$ values }
\label{alg:LBlambda} 
\algorithmicrequire{ $S_\lambda = \textrm{grid}\left(0.5:\lambda_{max}\right)$}\\
\algorithmicensure{ Estimated $\lambda_\text{best}$, $\hat{\boldsymbol{\beta}}_\text{first}$, $\hat{\boldsymbol{\beta}}_\text{best}$}\\
\vspace{-.5cm}
\begin{algorithmic}[1]
\small
\State $\hat{\boldsymbol{\beta}}_\text{first}, N_c \gets\textsc{LBOTDR}(\boldsymbol{y}, \lambda\!=\!0.5, \boldsymbol{\beta}_\text{start}\!=\!{\bf 0}, \boldsymbol{v}_\text{start} \!=\! {\bf 0}, N = N_\text{max})$
\State $\hat{\boldsymbol{\beta}} \gets \hat{\boldsymbol{\beta}}_\text{first}$
\State $b_\text{best} \gets \text{BIC}(\hat{\boldsymbol{\beta}}_\text{first}, \boldsymbol{y})$
\State $\hat{\boldsymbol{\beta}}_\text{best} \gets \hat{\boldsymbol{\beta}}_\text{first}$
\For{$\lambda \in S_\lambda$}
	\For {$j \in 1,\ldots, p$}
	\State $v_j = (|\beta_j| + \lambda) \text{sign} (\beta_j),$
	\EndFor
	\State $\hat{\boldsymbol{\beta}} \gets\!\textsc{LBOTDR}(\boldsymbol{y}, \lambda, \boldsymbol{\beta}_\text{start}\!\!=\!\boldsymbol{\beta}, \boldsymbol{v}_\text{start} \! \!=\! {\bf v}, N \! \!=\! 0.1 N_c)$
	\If{$\text{BIC}(\hat{\boldsymbol{\beta}}, \boldsymbol{y}) < b_\text{best}$}
		\State $b_\text{best} \gets \text{BIC}(\hat{\boldsymbol{\beta}}, \boldsymbol{y})$
		\State $\hat{\boldsymbol{\beta}}_\text{best} \gets \hat{\boldsymbol{\beta}}$
	\EndIf
\EndFor
\end{algorithmic}
\end{algorithm}

\section{Results}
\label{sect:results}

In order to provide an extensive analysis of the automatic fault location capabilities of the proposed LBI, an equally extensive testbench of fibers must be available. Since different fiber profiles with different combinations of events (fault magnitudes and positions) that could provide such an extensive testbench are not easily available, it is necessary to resort to simulation to enrich and broaden the analysis. Therefore, the result analysis is divided into two parts: first, the algorithm is tested against a testbench fiber profile, which has been manipulated to present challenging characteristics in terms of fault location; second, a methodology that is able to create simulated fiber profiles that mimic all the relevant characteristics of real-world profiles is presented. This way, an extensive testbench of simulated profiles based on this procedure is created, which, in turn, enables a statistically relevant analysis of the algorithm's performance.

\subsection{Experimental Results}

The complexity of a fiber profile may differ greatly, from a perfect fiber with no faults or events -- for instance, a recently installed fiber without any splitters --, to a fiber with multiple events of different magnitudes separated by arbitrary distances. The OTDR profile presented in Fig. \ref{fig:exampleOTDRTrace} is a good example of the latter kind of profile since it contains a variety of events: extremely high magnitude faults; low magnitude faults in noisy regions of the profile; and both low and high magnitude faults within a short spatial interval. This specific fiber, however, has been manipulated so that such features could be observed and, due to the challenge it poses in terms of precisely locating the events from a fault location algorithm perspective, has been used to attest the success of the proposed LBI algorithm in a pessimistic real-world environment. The monitoring method employed for the acquisition of this profile was the Wavelength Tunable Photon-Counting OTDR presented in detail in \cite{AmaralJLT2015}.

\begin{figure}[htbp]
\center
\includegraphics[width=0.9\linewidth]{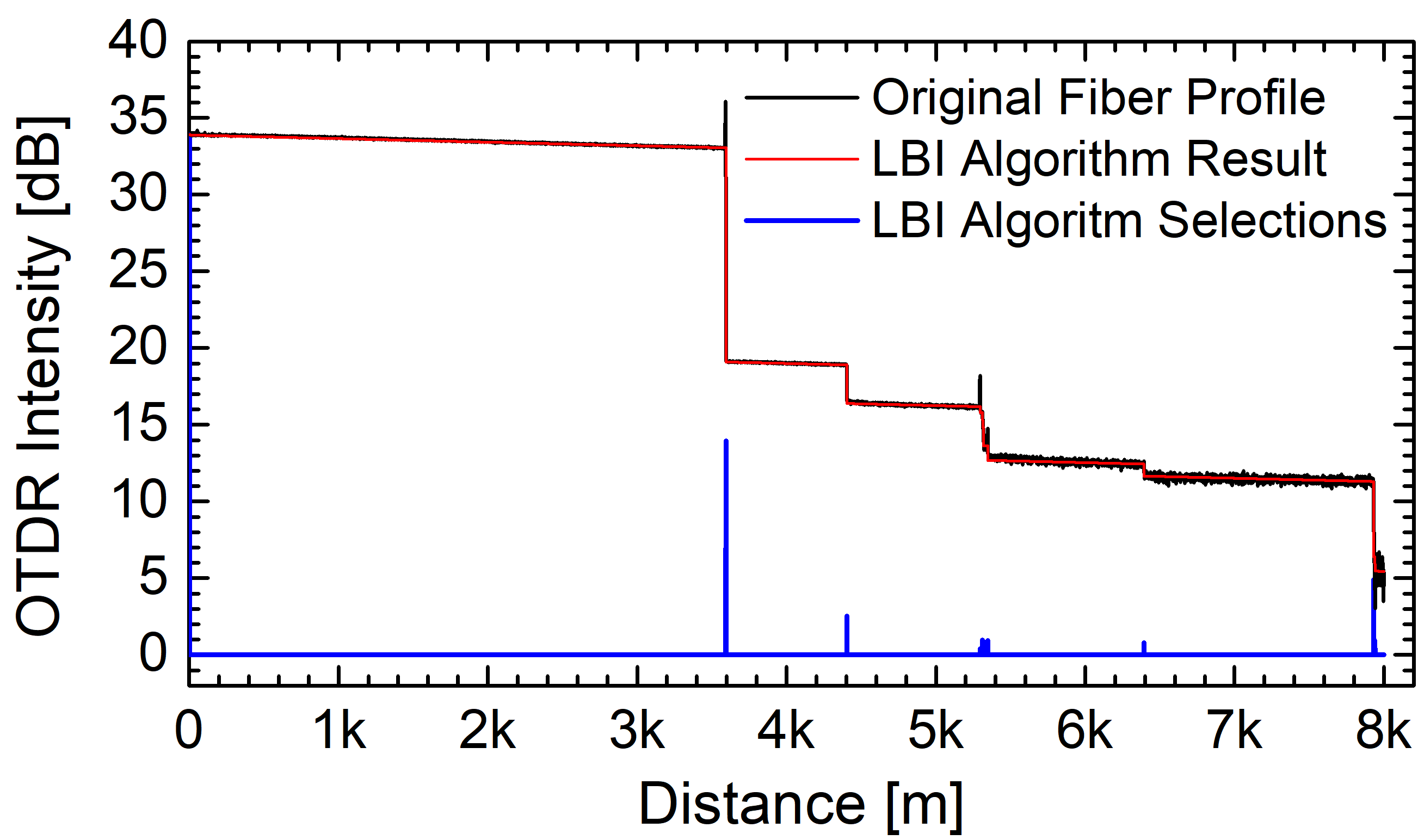}		
\caption{LBI Algorithm results for the testbench fiber profile. The original fiber profile is depicted in black, while the filtered result is presented in red. In blue, the vector $\hat{\boldsymbol{\beta}}_\text{best}$, that carries the information of position and magnitude of the selections of the algorithm, is depicted for reference.}
\label{fig:BregmanFilterResult}
\end{figure}

In Fig. \ref{fig:BregmanFilterResult}, the filtering results using the LBI algorithm are presented. Simultaneously, the comparison between positions (Real Pos.) and magnitudes (Real Mag.) of faults located by an operator that has access to the OTDR trace and the estimated positions (Est. Pos.) and magnitudes (Est. Mag.) retrieved by the algorithm is detailed in Table \ref{tab:expRes1}. It can be concluded that the estimated values are extremely close to the real values, with the maximum error in fault location being 2 meters and the maximum error in magnitude estimation being 0.2 dB. Therefore, even in a pessimistic environment, the proposed LBI algorithm excelled in determining the fault positions with both sensitivity and specificity, since no other events than the ones presented in Table \ref{tab:expRes1} were identified. The algorithm elapsed time for this profile was 150 seconds, and the chosen $\lambda$ value after the BIC model selection was $0.5$.

\begin{table}[htbp]
\centering
\caption{Event List Results for the Testbench Fiber Profile}
\begin{tabular}{c c | c c}
\textbf{Real Pos.} & \textbf{Est. Pos.} & \textbf{Real Mag.} & \textbf{Est. Mag.} \\
\hline
3593 m & 3594 m & 13.9 dB & 13.9 dB\Tstrut\\
4404 m & 4404 m & 2.4 dB & 2.5 dB\\
5298 m & 5297 m & 0.4 dB & 0.4 dB\\
5313 m & 5313 m & 1.1 dB & 1.0 dB\\
5317 m & 5319 m & 1.2 dB & 1.2 dB\\
5349 m & 5348 m & 0.6 dB & 0.9 dB\\
6395 m & 6395 m & 0.7 dB & 0.8 dB\\
7932 m & 7932 m & 5.8 dB & 5.6 dB
\end{tabular}
\label{tab:expRes1}
\end{table}

It interesting to evaluate the results of Fig. \ref{fig:BregmanFilterResult}: they not only depict an extremely close estimation of the filter result with respect to the original signal, but it also call the attention to the fact that, given a sparse $\boldsymbol{\beta}$ vector, a noiseless fiber profile can be simulated through ${\bf A} \boldsymbol{\beta}$. Indeed, the only striking difference between the red curve ($\bf{A\boldsymbol{\beta}}$) and the black curve ($\bf{y}$) of Fig. \ref{fig:BregmanFilterResult} is the absence of the noise component in the former. Thus, if one chooses an arbitrary vector $\boldsymbol{\beta}_\text{sim}$, a simulated noiseless fiber profile $\bf{y_\text{sim}}$ can be created through $\bf{y_\text{sim} = A \boldsymbol{\beta}_\text{sim}}$. The noise parcel must, therefore, be included in the model of the simulated noiseless fiber profile $\bf{y_\text{sim} = A\boldsymbol{\beta}_\text{sim}}$ so that the simulation approaches the real measurement. The main sources of noise in an OTDR profile are the uncertainty of data acquisition \cite{WeidJLT2016}, which can be modelled as a Poisson random variable, and the intrinsic coherent noise of coherent OTDR measurements \cite{de2006significance}, which can be modelled as an additive Gaussian random variable.

The uncertainty on data acquisition can be interpreted as the uncertainty of the number of photons measured by the photodetector. This counting process scales with the square root of the number of counts and reflects the lost of precision in the measurement as the signal power is reduced. This effect can be observed in positions of the fiber profile that are either far from the OTDR probe input or are positioned after drastic power drops -- as can be seen e.g. in Fig. \ref{fig:BregmanFilterResult}. Given a position $X$ in $\bf{y_\text{sim}}$, with $C$ counts, the counting noise is given by $\sqrt{C}$, and the signal-to-noise ratio (SNR) can be calculated as $\tfrac{C}{\sqrt{C}}$. Using the well-known relation $\text{SNR} = \tfrac{P_s}{P_n}$, ($P_s$ and $P_n$ being the signal and noise power, respectively) the counting noise variance for each position can be calculated and included in the simulated profile.

The noise originated from the CRN, on the other hand, can be modelled as an additive white Gaussian noise with variance calculated as given by \cite{de2006significance}:
\begin{equation}
\sigma_{CRN} =\left(\frac{V_g}{4\Delta z \Delta \nu}\right)^{1/2},
\end{equation}
where ${\bf v}_g$ is the group velocity inside the fiber (assumed to be $2\times10^8$ m/s with an index of refraction approximately equal to $1.5$), $\Delta z$ is the fiber length, and $\Delta \nu$ is the linewidth of the optical probing source. If one considers a probing source with spectral bandwidth as wide as the DWDM channel width ($\Delta \lambda = 0.8$ nm), it translates into $\Delta \nu = 100$GHz as specified in \textbf{ITUT G.694.1} \cite{ITU2012}, which causes the contribution of the CRN to be feeble. Since one of the goals of this study is to stress the fault-finding capabilities of the methodologies in noisy environments, a narrow linewidth probing source with $\Delta \nu=100$ kHz has been considered throughout the numerical simulations, which increases the contribution of the CRN.

\begin{figure}[htbp]
\center
\includegraphics[width=0.95\linewidth]{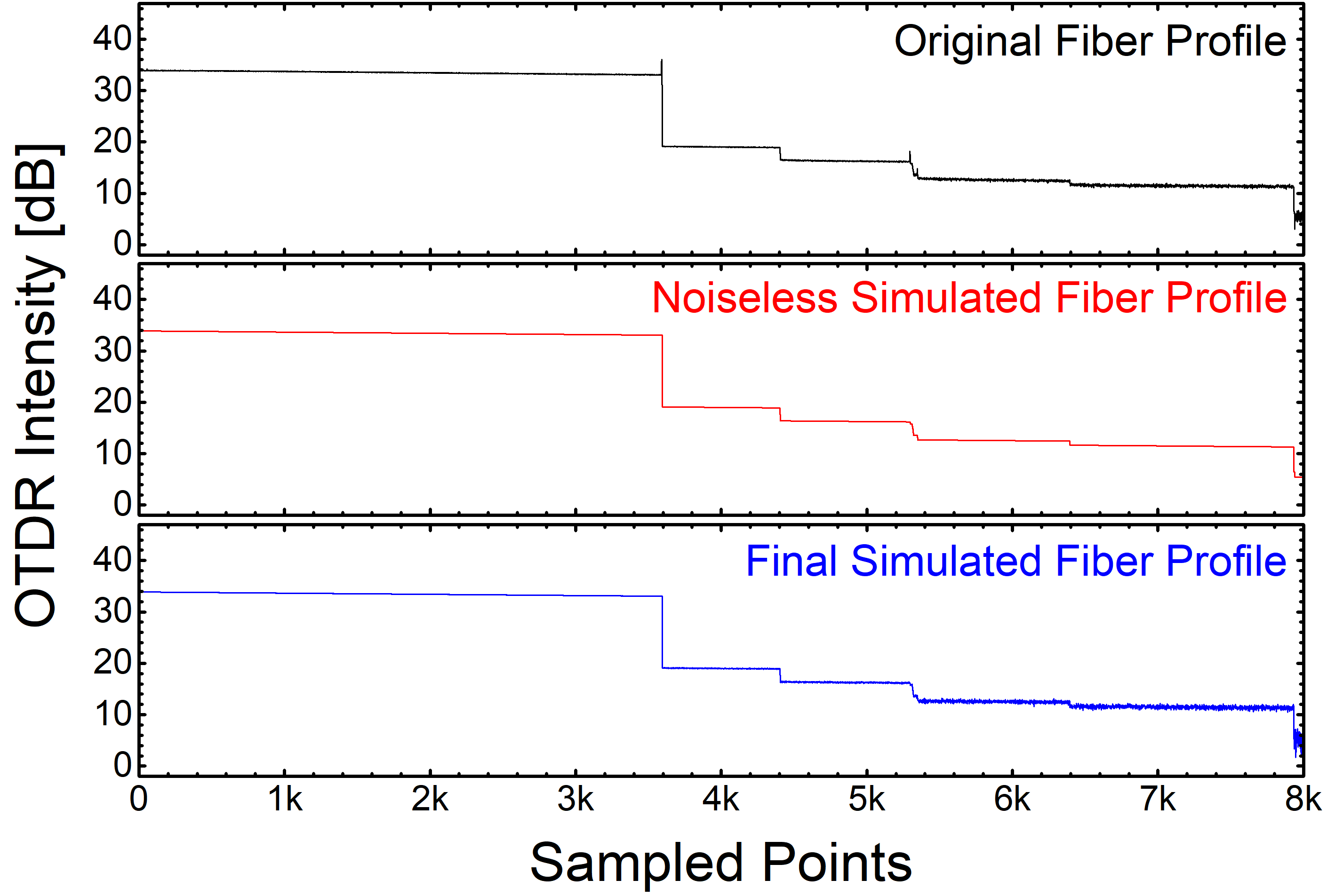}		
\caption{Generation of a simulated profile that exhibits the characteristics of real-world fiber profiles. In the first panel, the original profile of Fig. \ref{fig:BregmanFilterResult} is depicted as a reference, where the noise components can be clearly identified. In the second panel, the noiseless reconstructed fiber profile is presented, where the $\hat{\boldsymbol{\beta}}_\text{best}$ resulting from the processing of the original profile has been used. Finally, in the third panel, the simulated profile after inclusion of the noise components is presented.}
\label{fig:simExample}
\end{figure}

An example of a simulated fiber profile $\bf{y_\text{sim}}$ is depicted in the third panel of Fig. \ref{fig:simExample}. In order to illustrate the recovery of all the characteristics expected in a fiber profile by the simulation, $\boldsymbol{\beta}_\text{sim}$ was chosen to be equal to the estimated $\hat{\boldsymbol{\beta}}_\text{best}$ of the LBI algorithm after processing the testbench fiber profile of Fig. \ref{fig:BregmanFilterResult}. The three panels of Fig. \ref{fig:simExample} depict: the original testbench fiber profile (top); the noiseless $\bf{y_\text{sim}}$ constructed from the resulting $\boldsymbol{\beta}$ after processing of this profile (middle); and the final $\bf{y_\text{sim}}$ after introduction of the previously described noise components (bottom). In summary, the simulated fiber profiles used throughout the simulated testbench were generated by creating random sparse vectors $\boldsymbol{\beta_\text{sim}}$ and applying matrix $\bf{A}$ to generate controlled fiber profiles $\bf{y_\text{sim}}$ to which noise is included such that it simulates real-world fiber profiles. A striking resemblance to the original profile can be observed (first and third panels of Fig. \ref{fig:simExample}), validating the method for simulating fiber profile.

A comment should be made regarding the reflection peaks that are often present in the original OTDR measurement at fault positions but have not been reconstructed in the simulated profile. Even though the detection of such peaks could be included in the analysis algorithm \cite{souto2016l}, the main interest is in the detection of the trend breaks which indicate the presence of faults. These peaks, as far as the LBI algorithm is concerned, are treated as outliers, when not included in the model, and do not compromise its performance. In order to focus on the trend break detection, the inclusion of such peak signatures was not considered throughout the construction of the simulated profiles.

\subsection{Controlled Test Results}
The simulated testbench was created by randomly selecting positions and magnitudes of faults in profiles with sampled data points spanning from 5000 to 15000. On a standard OTDR device, and assuming the maximum dynamic range achievable, a profile with 15000 points corresponds to 100km, which is a sensible distance for long-haul optical networks and was thus chosen as the maximum number of points in the testbench. On the other hand, shorter fibers, within the tens of kilometers range, must also be contemplated, so 5000 sampled data points is set as the minimum number of points. Steps of 1000 data points from the minimum to the maximum number of data points are taken in order to provide a detailed dependence of the algorithm on the size of the original data. For each length, 1000 profiles were randomly selected and 5 faults of magnitudes out of a uniform distribution from 0.5 to 5 dB were randomly placed in the original data stream.

In order to provide a comparative analysis, the profiles have been analysed with the LBI Algorithm (Algorithm \ref{alg:LBlambda}) and with the Adaptive $\ell_1$ Filter, where the latter is taken as a reference. Gauging the success of the algorithms involves three parameters, as previously mentioned and informally set forward by \cite{WeidJLT2016}:
\begin{itemize}
\item sensitivity, the capacity to find the positions which correspond to faults.
\item specificity, the capacity to neglect all the positions which do not correspond to faults.
\item celerity, the capacity to be both sensitive and specific in the least amount of time. 
\end{itemize}
An analysis based on the computational complexity will also be drawn in the next section, since it is of the utmost importance when adapting the algorithm to an embedded processing unit, one of the sought for goals of automatic fault finding \cite{calliari2018high}.

\begin{figure}[htbp]
\center
\includegraphics[width=0.9\linewidth]{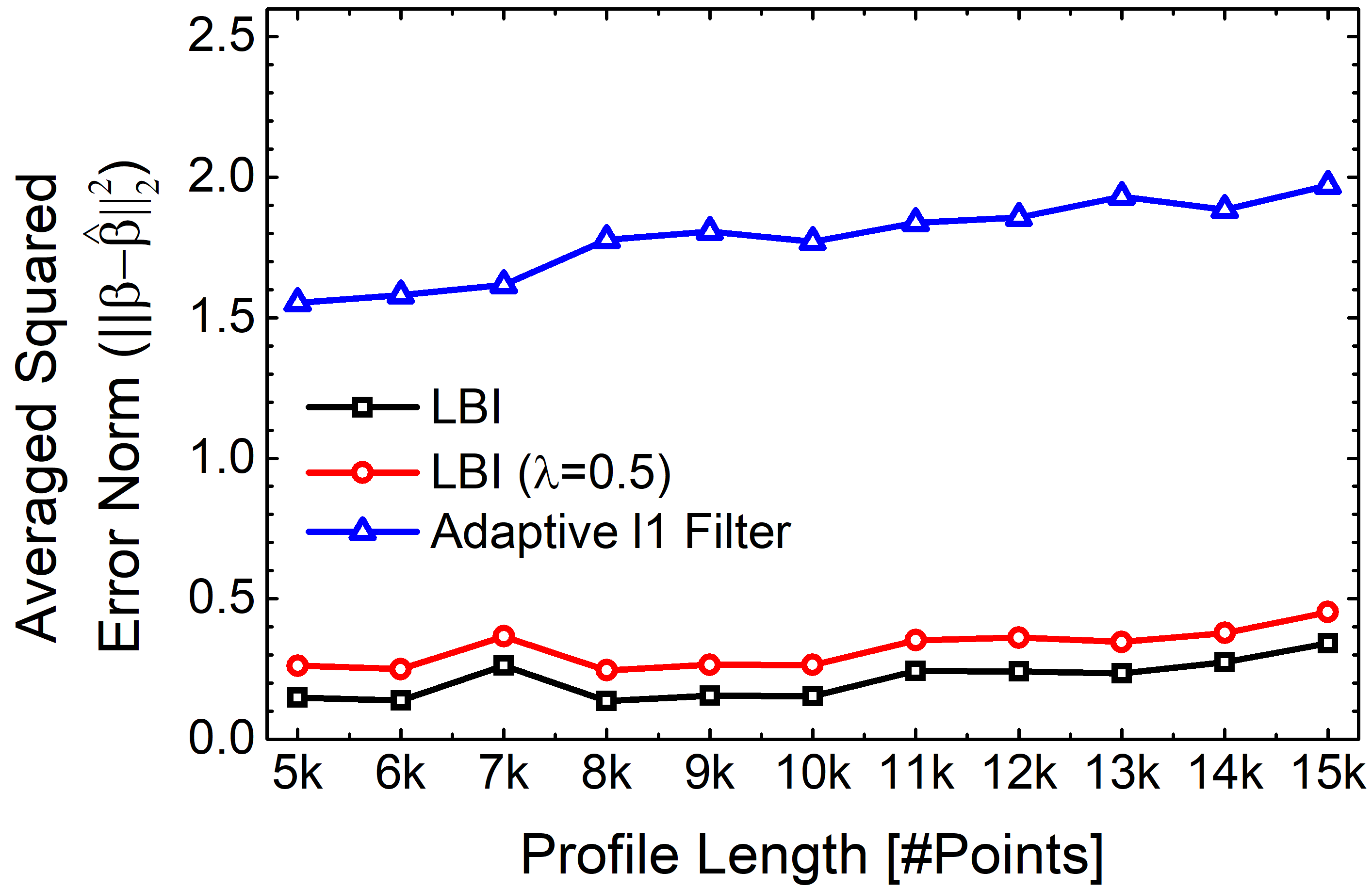}		
\caption{Averaged squared error norm between the recovered sparse vector $\hat{\boldsymbol{\beta}}$ and the original sparse vector $\boldsymbol{\beta}$ ($||\boldsymbol{\beta}-\hat{\boldsymbol{\beta}}||_2^2$). The average is between the 1000 different profiles randomly sorted for each of the data sizes. While the black curve correspond to the final output of the LBI algorithm, the red curve corresponds to the result of the first run, with $\lambda=0.5$.}
\label{fig:fibLenErr}
\end{figure}

Instead of analysing the results directly in light of the figures of merit described in the previous paragraph, a visually interesting and intuitive analysis is performed and presented in Fig. \ref{fig:fibLenErr}. That is the squared error norm between the original $\boldsymbol{\beta}$ used to create the fiber profile and the estimated coefficient vector $\hat{\boldsymbol{\beta}}$ averaged over all the random profiles for a given data size. This metric is useful since it considers errors in the estimation of the magnitudes of the faults as well as in the estimation of their positions. Since $\boldsymbol{\beta}$ is a sparse vector, lack of both sensitivity and specificity by the algorithms are easily identifiable from a deviation from the perfect zero-error mark of $||\boldsymbol{\beta}-\hat{\boldsymbol{\beta}}||_2^2$. Additionally, since the noise components of each fiber profile were simulated to mimic the practical reality, longer fibers will often exhibit positions with lower SNR than shorter fibers. Therefore, the slight increase in the averaged squared error norm observed when the data size increases is expected.

\begin{table*}[htbp]
\centering
\caption{Contingency Table -- Fault Finding}
\begin{tabular}{c c| c| c c| c|c}
 & \multicolumn{3}{c}{Fault Present} & \multicolumn{3}{c}{Fault Absent} \\
\cmidrule(lr){2-7}
Fault     & TP ($\ell_1$) & TP (LB) & TP (LB $\lambda=0.5$) & FP ($\ell_1$) & FP (LB) & FP (LB $\lambda=0.5$) \\
Found     & 75,389 & 74,814 & 74,811 & 777,800 & 30525 & 56808 \\
\cmidrule(lr){2-4} \cmidrule(lr){5-7}
Fault     & FN ($\ell_1$) & FN (LB) & FN (LB $\lambda=0.5$) & TN ($\ell_1$) & TN (LB) & TN (LB $\lambda=0.5$) \\
Neglected & 1599 & 2174 & 2177 & 109,145,212 & 109,892,487 & 109,866,204 \\
\cmidrule(lr){2-7}
Measures  & Sens. ($\ell_1$) & Sens. (LB) & Sens. (LB $\lambda=0.5$) & Spec. ($\ell_1$) & Spec. (LB) & Spec. (LB $\lambda=0.5$) \\
          & 97.92\% & 97.18\% & 97.17\% & 99.29\% & 99.97\% & 99.95\% \\     
\end{tabular}
\label{tab:ContTable}
\end{table*}

\begin{figure}[htbp]
\center
\includegraphics[width=0.9\linewidth]{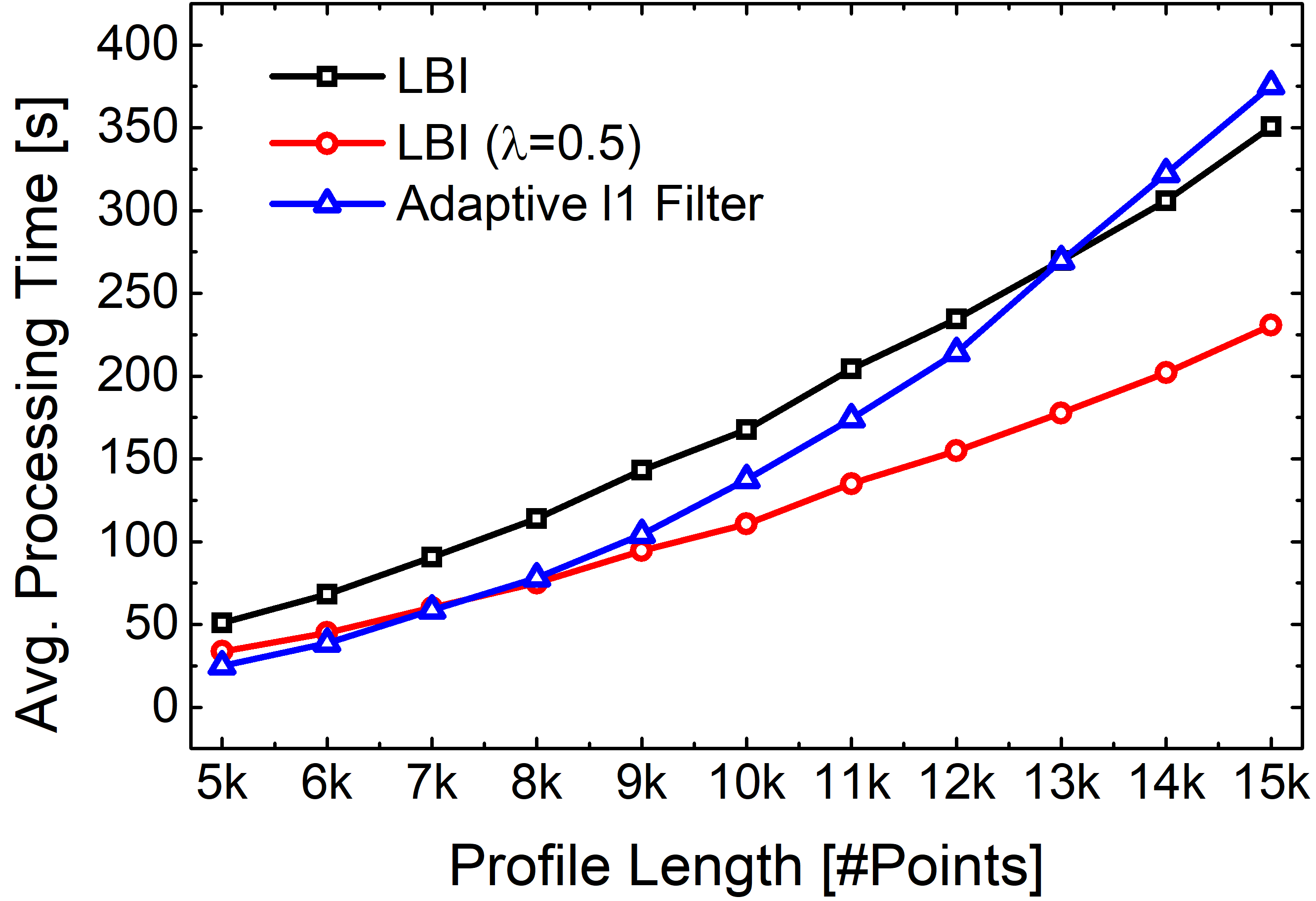}		
\caption{Algorithm elapsed time averaged over the 1000 different profiles randomly sorted for each profile length. For long profiles, the Linearized Bregman Iterations reach convergence faster than the Adaptive $\ell_1$ Filter.}
\label{fig:fibLenTime}
\end{figure}

In Fig. \ref{fig:fibLenErr}, the results considering the output of the proposed LBI algorithm (black) are depicted along with those after the first run (red), with $\lambda=0.5$. This attests that, as commented in Section \ref{sect:3}, running the algorithm based on the stopping criteria and $\lambda=0.5$ yields high-quality results. Also, as can be observed in Fig. \ref{fig:fibLenTime}, the total average processing time increases substantially when the search for the best $\lambda$ is performed while the gain in accuracy is not extreme. Therefore, an operator facing the compromise between longer processing times and slight accuracy gains may opt to stop Algorithm \ref{alg:LBlambda} in line 4. Fig. \ref{fig:fibLenTime} also shows that the average processing time for both algorithms is comparable for the range of sampled data points considered in this testbench. It is clear, however, that the slope of the blue curve increases faster, indicating that the Adaptive $\ell_1$ Filter algorithm performs slower than the proposed LBI algorithm for bigger data sets.

The trends presented in Figs. \ref{fig:fibLenErr} and \ref{fig:fibLenTime} indicate that both algorithms were successful in recovering the original sparse vector used to create the simulated fiber profiles in the presence of noise while maintaining comparable processing times. Fig. \ref{fig:fibLenErr} in particular, however, shows a clear dominance of the LBI filter over the Adaptive $\ell_1$ Filter with respect to the proximity of the estimated to the original profile. This is qualitatively translated into an error more than $5$ times smaller even when disregarding the $\lambda$ selection step of Algorithm \ref{alg:LBlambda}.

Even though the results of Fig. \ref{fig:fibLenErr} are clear, a means of assessing the figures of merit of sensitivity and specificity and analyse the performance of the algorithms with greater precision is desirable. In this context, contingency tables are specially interesting for the analysis of binary classification tests. Additionally, the outputs of a contingency table analysis are exactly the figures of merit of interest. By interpreting the trend break detection problem as a \textit{break/no break} -- and, thus, a binary --- classification test, the contingency table for the results of the algorithms can be constructed. In Table \ref{tab:ContTable}, the absolute values of True Positives (TP), False Positives (FP), True Negatives (TN), and False Negatives (FN) are presented for the LBI and for the Adaptive $\ell_1$ Filter Algorithms.

To perform this analysis, the entries of the contingency table were defined as follows: the TPs correspond to faults found by the algorithms which are present in the original sparse vector; the TNs correspond to positions that were not selected by the algorithms and that did not originally contain faults; the FPs are faults found by the algorithms in positions where there were originally no faults and the FNs correspond to positions not selected by the algorithm but that originally contained faults. Throughout the analysis, matches were considered with a zero error margin, i.e., matches are only considered if the estimated position is exactly the same as the original position. Also, for the construction of this contingency table, the results from all the simulation runs have been taken into account without discrimination with respect to data size, as in Fig. \ref{fig:fibLenErr}.

Regarding the Sensitivity and Specificity, both algorithms exhibit extremely high values for both figures of merit. The \textit{Accuracy}, defined as $\tfrac{TP+TN}{TP+FP+FN+TN}$, is also extremely high with values of 99.97\% and 99.29\% for the proposed LBI algorithm and the Adaptive $\ell_1$ Filter, respectively, indicating that both algorithms are extremely well suited for the ascribed trend break detection problem. A closer analysis, however, shows that the small differences in Accuracy may have a deeper meaning which is clouded by the high number of TNs, so it is useful to compare the absolute numbers of FPs and FNs. To put the absolute numbers in perspective, an overall number of $110$ Million potential fault positions have been simulated altogether for the presented results. 

While the LBI failed to identify 575 more faults than the Adaptive $\ell_1$ Filter, which corresponds to an increase of $\sim35\%$ in the FNs, the Adaptive $\ell_1$ Filter wrongly identified 747,275 more points as faults than the LBI, an increase of $\sim2,400\%$ in the FPs. This analysis shows that, even though the Adaptive $\ell_1$ Filter is slightly less likely to neglect a trend break that is present in the data stream, it comes at the expense of flooding the result with several candidates which do not correspond to actual trend breaks. At the same time, the equilibrium between FPs and FNs seems to be more closely achieved by the LBI algorithm. Additionally, the magnitudes of the faults neglected by the LBI were studied, showing that the algorithm fails to identify small faults, i.e., those that correspond to a 0.5 dB power drop in the profile -- the minimum value used in the simulation procedure. This indicates that the LBI is more sensitive and specific to faults that can indeed pose threats to the link operation.

\begin{figure}[htbp]
\center
\includegraphics[width=0.9\linewidth]{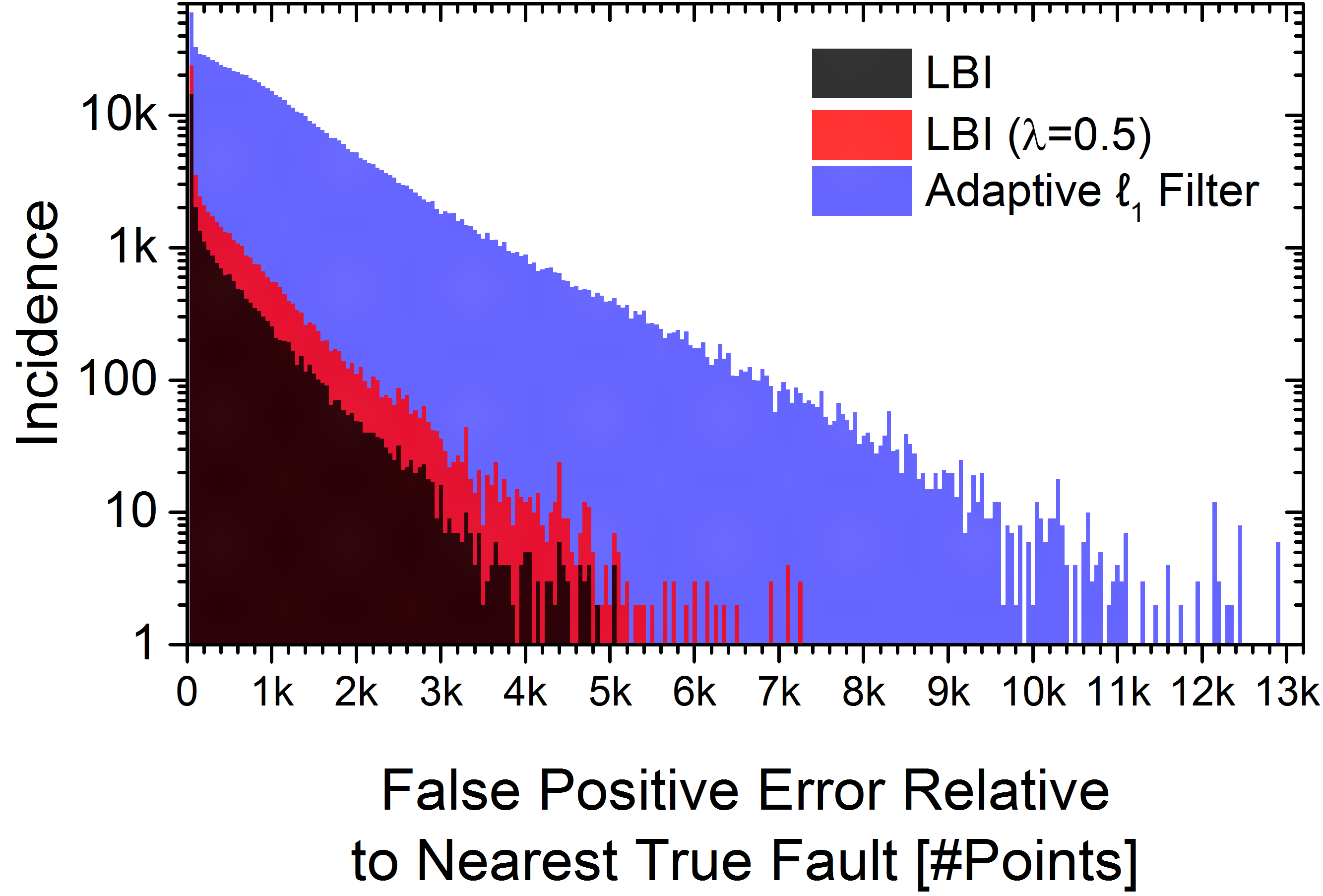}
\caption{Histogram of False Positive incidence with respect to the distance from nearest real fault. The histogram bin width is 50 points and the logarithmic scale is used for clearer visualization of the results.}
\label{fig:FPHist}
\end{figure}

To get a better impression on the relative error distribution of the results summarized in the FP (False Positive) class, the histogram of FP incidence with respect to its distance from the nearest real fault is depicted in Fig.~\ref{fig:FPHist} for both Algorithm 2 (with and without the selection of $\lambda$'s) and the Adaptive $\ell_1$ Filter. For this analysis, the following evaluation was employed: for each processed simulated fiber profile, the minimum absolute value difference between entries of the output vector $\hat{\boldsymbol{\beta}}$ and the original vector $\boldsymbol{\beta}$ were determined, i.e., the False Positive's distance to the nearest true fault position. 

The histograms clearly exhibit different widths, translating the overall proximity of a False Positive result to a real fault. It is interesting to note that the widths of the LBI and LBI ($\lambda = 0.5$) histograms already differ, indicating that the impact of the $\lambda$ selection stage is majoritarily on the FPs, as already evidenced by the Contingency Table. The most striking result of Fig. \ref{fig:FPHist}, however, is the major width difference between the LBI algorithm and the Adaptive $\ell_1$ Filter histograms, corroborating the previous analysis that the latter, even though having a higher TP rate, floods the selection result with False Positives. In this context, the \textit{Precision} figure of merit, defined as $\tfrac{TP}{TP+FP}$, is relevant as it translates the confidence of an operator on a positive result output from the analysis algorithm. Even with a lower number of TPs, the confidence on a positive result for the LBI ($\sim71\%$) is much higher than for the Adaptive $\ell_1$ Filter ($\sim8\%$).

\color{black}
\subsection{Complexity}
\label{sect:cmplx}
The complexity analysis for the Adaptive $\ell_1$ Filter has been presented in \cite{WeidJLT2016}. There, the authors state that ``\textit{a model that can be described by $m$ components will take $O\left(pm\right)$ operations}" [per iteration], where $p$, in this document, stands for the number of rows of the matrix \textbf{A}. The number $m$ refers to the number of non-zero elements in a current estimate of $\boldsymbol{\beta}$ during the iterations. The number of iterations for the Adaptive $\ell_1$ Filter is not preselected, the algorithm typically runs until convergence, i.e., until the non-zero positions do not change between two consecutive iterations.

The presented Algorithm 2 on the other hand only requires $O(N)$ operations per iteration. More specifically, as can be seen in Algorithm 2, its main processing steps (lines $9$ and $10$ of Algorithm 2) only require $2$ multiplications and $i+2$ additions at iteration $k$ with $i = (k-1) \text{ mod } p +1 \in {1,\ldots, p}$). On average this means that this step only requires $2$ multiplications (the value of $1/\|{\bf a}_i\|_2^2$ can be pre-calculated and stored in memory) and approximately $p/2$ additions. Based on the same considerations one can see that the lines $11$ to $13$ add a single multiplication and approximately another $p/2$ additions per iteration on average. Combining this with the $p$ subtractions of the shrink operation, this means that a single iteration of the Algorithm can be performed with about $2p$ addition-equivalents on average, as well as with $3$ multiplications. 

The Adaptive $\ell_1$ Filter is typically calculated multiple times for a set of $\lambda$ values, usually processed in descending order. This allows using the results of a previous calculation using a higher $\lambda$ value to warm-start a following calculation with a lower value. This is a reasonable approach for an active set-type algorithm, where a large value of $\lambda$, due to the high sparsity of the expected result, leads to a fast convergence of such algorithms. Starting the Adaptive $\ell_1$ Filter with a small $\lambda$ value would result in low sparsity, i.e. a huge active set, leading to a high computational complexity due to a large number of required iterations. 

For Linearized Bregman-based algorithms, an opposite behaviour can be observed. In terms of convergence speed, $\lambda$ should be chosen as small as possible. This is because, although the complexity of a single LB Iteration does not depend on the sparsity of the result, using higher $\lambda$ values typically increases the number of required iterations until convergence. Useful values of $\lambda$ are generally within the range of $[0, \lambda_\text{max})$, where $\lambda_\text{max}$ is the value where the zero vector is the solution of (\ref{eq:linearizedBregman}). For $\lambda_\text{max}$ one can typically calculate upper bounds (e.g. as described in \cite{Friedman, lambdamax}). However, as evaluated in this section, successful sparse estimation of trend breaks in OTDR profiles can be achieved with values close to $0$, giving LB based algorithms an additional advantage in terms of computational complexity over the Adaptive $\ell_1$ Filter.

Finally, as also stated in \cite{WeidJLT2016}, the Adaptive $\ell_1$ Filter requires a normalization of the columns of the $\bf A$ matrix to achieve a good performance. Such normalization requires additional multiplications compared to when using $\bf A$ directly, where columns with only ones and zeros enable performing scalar products with additions. Compared to the Adaptive $\ell_1$ Filter, the LB algorithm can use the matrix $\bf A$ directly without any pre-processing.

\subsection{Digital Hardware Implementation and Improvements on the LBI}

When analysing Algorithm 2, one can see that this algorithm is perfectly suited for implementation in digital hardware: the main computational complexity is spent on parallel additions. When considering an implementation on an FPGA, by storing the vectors $\boldsymbol{\beta}$ in parallel block RAMs, one can access the values of the vectors in parallel. Providing such an access for hundreds of values in parallel poses no problem to even medium level FPGAs, e.g., as described in \cite{cyclon4deviceov} for 
Altera Cyclon IV FPGAs. Adding up a large number of values in parallel can also be easily performed by using pipelined adder trees \cite{cookbook}. This allows for a high level of parallelization for the LB based algorithm enabling an efficient development of an unified measurement and analysis hardware in a single embedded unit.

For LB based sparse Kaczmarz, recently a variety of improvements have been published such as efficient Linearized Bregman Iterations \cite{lunglmayr2016efficient}, mikrokicking \cite{CAMSAP}, or using a sparsity-adaptive step width  \cite{SSP}. Such approaches significantly allow reducing the complexity of LB based Sparse Kaczmarz algorithms. To provide a general comparison on an algorithmic level, it was decided not to include such enhancements in this work. However, each of the mentioned enhancements can be easily applied to the LB iterations for OTDR analysis, allowing reducing the complexity even further.

\section{Conclusion}
The increasing demand for high-rate communication links can be met with new generation optical fiber networks. An important aspect of such networks, for both the user and the network operator, is robustness, which translates into shorter downtimes for the users and reduced operational costs for the operator. In this work, a measurement and processing system for fault analysis and detection for fiber optic links is presented, using a Linearized Bregman Iterations based algorithm especially designed for this problem. In-depth performance evaluation has been presented to support the claims.

Application of the Linearized Bregman Iterations method by means of a tailored algorithm has been successful in real and simulated fiber profiles. By conjugating a simple procedure to generate arbitrary noiseless profiles and an analysis of the main noise contributions that affect the data stream permitted a close reconstruction of fiber profiles that share the same characteristics as found in practice. With such a methodology at hand, an extensive testbench was created, where fibers with distinct characteristics could be analyzed by the proposed algorithm and a reference trend break detection algorithm found in the literature. Analysis of the results was performed evidencing the suitability of the algorithms and highlighting the dominance of the LBI algorithm with respect to the equilibrium between sensitivity and specificity with high accuracy. 

A timing comparison indicated that LBI based algorithms converge faster for long fiber profiles. In addition, a smoother slope was observed in the processing time of the proposed algorithm as the instance size increase, hinting to the fact that it exhibits reduced complexity. To further investigate this point, a complexity analysis of both algorithms was conducted confirming the lower complexity of the LBI based algorithm. Finally, a brief rationale points toward the possibility of embedding the algorithm in an FPGA with expected results such as: significant reduction of processing time due to the parallelizability in dedicated hardware; and the development of an unified measurement and analysis hardware in a single embedded unit.

In conclusion, the results show that the novel algorithm outperforms the comparison algorithm in terms of estimation performance as well as computation speed, while still maintaining the algorithmic structure that can be efficiently implemented in digital hardware. Therefore, the proposed measurement system consisting of a Photon-Couting OTDR as data acquisition subsystem and the LBI algorithm as the data processing unit enables reliable automatic fault detection for optical fiber links.

\section*{Acknowledgment}
The authors would like to thank F. Calliari, J. D. Garcia, and M. Souto for their assistance with the Julia programming. G. C. Amaral is indebted to J. D. Garcia for enlightening discussions. Financial support from brazilian agencies CAPES, FAPERJ, and CNPq are acknowledged by G. C. Amaral.
This work has been supported by the COMET-K2 ``Center for Symbiotic Mechatronics'' of the Linz Center of Mechatronics (LCM) funded by the Austrian federal government and the federal state of Upper Austria.

\vspace{0.82cm}
\bibliographystyle{IEEEtran}
\bibliography{References}



%

\end{document}